\documentclass[12pt, epsfig]{article}
\usepackage{color}
\usepackage{graphicx}
  \usepackage{amsthm,amsfonts}
  \usepackage{amsmath}
\newcommand{\bea}   {\begin{eqnarray}}
\newcommand{\eea}   {\end{eqnarray}}
\begin{document}
\renewcommand{\thefootnote}{\fnsymbol{footnote}}

\thispagestyle{empty}

\title{On Light-like Deformations of the Poincar\'e Algebra}

\author{Zhanna Kuznetsova\thanks{{\em e-mail: zhanna.kuznetsova@ufabc.edu.br}}
~and~ Francesco
Toppan\thanks{{\em e-mail: toppan@cbpf.br}}
\\
\\
}
\maketitle

\centerline{$^{\ast}$
{\it UFABC, Av. dos Estados 5001, Bangu,}}{\centerline {\it\quad
cep 09210-580, Santo Andr\'e (SP), Brazil.}
\centerline{$^{\dag}$
{\it CBPF, Rua Dr. Xavier Sigaud 150, Urca,}}{\centerline {\it\quad
cep 22290-180, Rio de Janeiro (RJ), Brazil.}
~\\
\maketitle
\begin{abstract}

We investigate the observational consequences of the light-like deformations of the Poincar\'e algebra induced by the jordanian
and the extended jordanian classes of Drinfel'd twists. Twist-deformed generators belonging to a Universal Enveloping Algebra
close nonlinear algebras. In some cases the nonlinear algebra is responsible for the existence of bounded domains of the deformed generators. The Hopf algebra coproduct
implies associative nonlinear additivity of the multi-particle states.
A subalgebra of twist-deformed observables is recovered whenever the twist-deformed generators are either
hermitian or pseudo-hermitian with respect to a common invertible hermitian operator.
~\\
\end{abstract}
\vfill

\rightline{CBPF-NF-002/18}

\newpage

\section{Introduction}

This paper addresses the problem of the observational consequences of twist-deforming the
Poincar\'e algebra.  We work within a quantization scheme which has been previously applied to Drinfel'd twist deformations of quantum theories in a non-relativistic setting  .  Open questions are investigated. We mention, in particular, the nature of the observables: which of them are consistently maintained in the deformed theory either as hermitian or pseudo-hermitian operators?  To be specific, in the large class of deformed Poincar\'e
theories (which include, e.g., $\kappa$-Poincar\'e theories \cite{kpoi1}-\cite{kpoi3}, Deformed Special Relativity theories 
\cite{{dsr1},{dsr2}}, light-like noncommutativity in Very Special Relativity \cite{{jab1},{jab2}}  and many other examples \cite{ms1}-\cite{ncgra}) we focus on
the Drinfel'd twist \cite{{dri},{res}} deformations of a light-like direction.  Due to this reason the deformations we consider here are based on the jordanian \cite{dvl}-\cite{ogi}
and on the extended jordanian \cite{kul} twist (for physical applications of the Jordanian twist see \cite{{blt},{bopa1}} and, for the extended Jordanian twist, \cite{bopa2}-\cite{m2p2}).\par
The deformations {\textcolor{black}{(see Appendix {\bf B} for a more detailed discussion)}  can be encoded in twist-deformed generators which, essentially, correspond to 
 twist-covariant generalizations of the Bopp-shift \cite{bop}. The operators which in the undeformed case are associated with generators of the Lie algebra are, under a twist, mapped into given elements of the Universal Enveloping Algebra.\par
Observables in connection with deformed generators were addressed in \cite{abp} for twist deformations of the Poincar\'e-Weyl algebra with the dilatation operator entering the twist.  In our work the twist is defined in terms of Poincar\'e generators only. {\textcolor{black}{The framework we are using in this paper is based on \cite{ourtwist1}-\cite{ourtwist4}. Its presentation is quickly summarized in Appendix {\bf B}.
}}
 \par
Different deformed theories are obtained from the original twist and its flipped version (obtained by a permutation of the tensor space).\par
It is worth mentioning that quantum deformations of the Poincar\'e algebra were classified in \cite{zak} in terms of classical $r$-matrices for Poincar\'e Poisson structures. The connection with twist-deformations was presented in \cite{tol}. The extended Jordanian twist discussed below is recovered from the second case of Table $1$ given in \cite{tol} by setting equal to zero two of the three deformation parameters. 
\par
A common feature of the deformed theories is that the deformed generators define a closed nonlinear algebra
(antisymmetry and Jacobi identities are respected, but the right hand side is a nonlinear combination of the generators).
Furthermore, the deformation modifies the domain of the physical parameters. For instance, in the simplest non-trivial case, the jordanian deformation implies the introduction of a maximal momentum along a light-cone direction.
Induced by the coproduct (see \cite{ourtwist4}), nonlinear addition formulas are obtained for multi-particle states. The nonlinear addition formulas satisfy the associativity condition. Their main {\em raison d'\^{e}tre} is that they allow to respect the domain of validity of the physical quantities (in the example above, the composite momentum along the light-cone direction is bounded by the maximal value). We postpone to the Conclusions a more detailed discussion of the implications of our results.\par

 The scheme of the paper is as follows. The jordanian and the extended jordanian twists are recalled in Section {\bf 2}.  In Section {\bf 3} twist-deformed generators are introduced. In Section {\bf 4} the (pseudo)-hermiticity property
of twist-deformed  generators is discussed.  The arising of a nonlinear algebra is investigated in Section {\bf 5}. The bounded domains of deformed physical observables and nonlinear additive formulas are discussed in Section {\bf 6}. For completeness in the Appendix {\bf A} the (undeformed) Poincar\'e algebra in the light-cone basis is presented
{\textcolor{black}{and in Appendix {\bf B} the framework of twist-deformation of Hopf algebra used in this paper is succinctly explained.}}

\section{Jordanian versus extended Jordanian twist}

We recall, see \cite{{abe},{maj}} for details, that a Drinfel'd twist deformation of a Hopf Algebra $A$ is induced by an
invertible  element ${\cal F} \in A\otimes A$ which satisfies the cocycle condition
\bea\label{cocyclecondition}
({\bf 1}\otimes {\cal F})(id\otimes \Delta){\cal F} &=& ({\cal F}\otimes {\bf 1})(\Delta \otimes id) {\cal F}.
\eea
In Sweedler's notation \cite{swe} ${\cal F}$ can be expanded according to
\bea
{\cal F} =f^\beta\otimes f_\beta &,&
{\cal F}^{-1} ={\overline f}^\beta\otimes {\overline f}_\beta.
\eea
For $A={\cal U}({\cal G})$, the Universal Enveloping Algebra of a Lie algebra ${\cal G}$, the elements of ${\cal F}$ are taken from an even-dimensional subalgebra of ${\cal G}$. Therefore the simplest cases of twist are found for two-dimensional subalgebras.\par
There are (over $\mathbb{C}$) two inequivalent two-dimensional Lie algebras (we denote the generators as $a,b$):\par
{\em i}) the abelian algebra $[a,b] =0$ and \par
{\em ii}) the non-abelian algebra $[a,b]=i b$.\par
The case {\em i}), the abelian twist, leads to constant non-commutativity (see, e.g., \cite{{asc},{ourtwist1}}).\par
In this paper we focus on the second case. 
The non-abelian algebra {\em ii}) is, for example, the Borel subalgebra of $sl(2)$. 
The $sl(2)$ generators can be presented as $D,H,K$, satisfying the commutation relations
\bea
&\relax [D, H] = iH,\quad [D, K]=-iK, \quad [K,H] = 2i D.&
\eea
We can identify $D\equiv a$ and $H\equiv b$. This leads, see \cite{ourtwist3}, to non-commutativity of Snyder type.\par
We can also regard {\em ii}) as the subalgebra of a $d$-dimensional Poincar\'e algebra ${\cal P}(d)$. From its generators $P_0,P_1, M_{01}$, whose commutators are
\bea
&\relax [P_0,P_1]= 0,\quad
\relax [M_{01}, P_0]=-iP_1,\quad
\relax [M_{01},P_1]= iP_0,&
\eea
one can identify the {\em ii}) subalgebra from the positions
$a\equiv -M_{01}$, $b\equiv P_+=P_0+P_1$.
\par
The abelian algebra {\em i}) induces the abelian twist
\bea
{\cal F} &=& \exp (-i \alpha a\otimes b),
\eea
where $\alpha$ is the (dimensional) deformation parameter.\par
The non-abelian algebra {\em ii}) induces the non-abelian (jordanian) twist \cite{dvl}-\cite{ogi}
\bea
{\cal F} &=& \exp (-i a\otimes \ln (1+ \alpha b)),
\eea
where $\alpha$ is the (dimensional) deformation parameter.\par
Under the transposition operator $\tau$ ($\tau (v\otimes w)= w\otimes v$), the transposed twist
\bea
{\cal F}_\tau &:=& \exp (-i\ln(1+\alpha b)\otimes a),
\eea
still satisfies the cocycle condition (\ref{cocyclecondition}). We are using both ${\cal F} $ and ${\cal F}_{\tau}$ in
our paper.\par
The jordanian and the extended jordanian twist of the $d$-dimensional Poincar\'e algebra can be expressed, in terms of light-cone coordinates (see the Appendix
for the $d=4$ case) and Einstein's convention, through the position (see, e.g., \cite{jkm})
\bea\label{unflippedtwist}
{\cal F} &=& \exp\left( i M\otimes \ln(1+\alpha P_+) +i\epsilon M_{+j}\otimes\ln(1+\alpha P_+)\frac{P_j}{P_+}\right).
\eea
The jordanian case is recovered for $\epsilon=0$; the extended jordanian case is recovered for $\epsilon=1$. In two dimensions the two twists coincide since there are no transverse directions.\par
Under transposition, the ${\cal F}_\tau$ twist is given by
\bea\label{flippedtwist}
{\cal F}_\tau &=& \exp\left(i\ln(1+\alpha P_+)\otimes M + i\epsilon \ln(1+\alpha P_+)\frac{P_j}{P_+}\otimes M_{+j}\right).
\eea
The following four kinds of twist-deformations can be considered: the jordanian deformations ($\epsilon=0$) based on ${\cal F}$ (case $I$) and ${\cal F}_\tau$ (case $II$) and the extended jordanian deformations
($\epsilon=1$) based on ${\cal F}$ (case $III$) and ${\cal F}_\tau$ (case $IV$).

\section{Twist-deformed generators}

A twist deformation can be expressed in terms of the twist-deformed generators (see \cite{{asc},{asc2},{ourtwist1}}). 
{\textcolor{black}{To make the paper self-consistent, the framework we are using is succinctly described in Appendix {\bf B}.}}
Under deformation, a Lie algebra generator $g\in{\cal G}$ is mapped into the Universal Enveloping Algebra element $g^{\cal F}\in{\cal U({\cal G})}$, given by
\bea\label{twistdefmap}
g^{\cal F} &=& {\overline f}^\beta(g) {\overline f}_\beta, \quad ({\cal F}^{-1}={\overline f}^\beta\otimes
{\overline f}_\beta).
\eea
In this paper we consider {\textcolor{black}{the Hopf algebras ${\cal U}({\cal G})$ and ${\cal U}({\cal P})$. They are defined over the Universal Enveloping Algebras of, respectively,  the Lie algebra ${\cal G}=\{P_\mu, M_{\mu\nu}, x_\mu, \hbar\}$
and its subalgebra ${\cal P}=\{P_\mu, M_{\mu\nu}\}$, which is the $d$-dimensional Poincar\'e algebra (see Appendix {\bf A} for the explicit presentation of ${\cal G}$ and ${\cal P}$ for the ordinary $d=4$ spacetime).}}\par
We point out that $\hbar$ is a central element of ${\cal G}$ and is a primitive element of the Hopf algebra structure defined on the Universal Enveloping Algebra ${\cal U}({\cal G})$
(for details and motivations of the construction see, e.g.,  \cite{ourtwist4}).\par
We present the twist deformations for the previous Section cases $I$, $II$ and $IV$, whose twist deformed generators can be presented in closed form.\par
In the case $I$ (${\cal F}$ twist with $\epsilon=0$) we have
\bea
P_+^{\cal F}&=& P_+\frac{1}{1+\alpha P_+} ,\nonumber\\
P_-^{\cal F}&=& P_-(1+\alpha P_+) ,\nonumber\\
P_j^{\cal F}&=& P_j ,\nonumber\\
M^{\cal F}&=& M,\nonumber\\
M_{+j}^{\cal F}&=&M_{+j}\frac{1}{1+\alpha P_+} ,\nonumber\\
M_{-j}^{\cal F}&=&M_{-j}(1+\alpha P_+) ,\nonumber\\
N^{\cal F}&=&N
\eea
and
\bea
\hbar^{\cal F} &=& \hbar,\nonumber\\
x_+^{\cal F}&=& x_+\frac{1}{1+\alpha P_+},\nonumber\\
x_-^{\cal F}&=&x_-(1+\alpha P_+),\nonumber\\
x_j^{\cal F}&=&x_j.
\eea
The undeformed generators can be expressed in terms of the deformed generators on the basis of inverse formulas. In particular we have
\bea
P_+ &=& P_+^{\cal F}\frac{1}{1-\alpha P_+^{\cal F}},\nonumber\\
\frac{1}{1+\alpha P_+}= 1-\alpha  P_+^{\cal F} &, & 1+\alpha P_+ = \frac{1}{1-\alpha P_+^{\cal F}}.
\eea
The twist-deformed generators for the transposed twists ${\cal F}_\tau$, with $\epsilon=0,1$, are
\bea\label{twistdefiione}
P_\bullet^{\cal F}&=& P_\bullet,\nonumber\\
M^{\cal F}&=& \frac{1+2\alpha P_+}{1+\alpha P_+}M+\epsilon\left(\frac{\alpha P_j}{1+\alpha P_+} -\ln(1+\alpha P_+)\frac{P_j}{P_+}\right)M,\nonumber\\
M_{+j}^{\cal F}&=&M_{+j}+\epsilon\ln(1+\alpha P_+)M_{+j} ,\nonumber\\
M_{-j}^{\cal F}&=&M_{-j}+\frac{2\alpha P_j}{1+\alpha P_+} M +\epsilon\left( 
\ln(1+\alpha P_+)\frac{P_j}{P_+}\delta_{jk} +\frac{2\alpha P_jP_k}{(1+\alpha P_+)P_+}-
2\ln(1+\alpha P_+)\frac{P_jP_k}{{P_+}^2}\right) M_{+k},\nonumber\\
N^{\cal F}&=&N-\epsilon\epsilon_{jk}\ln(1+\alpha P_+)\frac{P_k}{P_+}M_{+j}
\eea
and
\bea\label{twistdefiitwo}
\hbar^{\cal F} &=& \hbar,\nonumber\\
x_+^{\cal F}&=&x_+,\nonumber\\
x_-^{\cal F}&=&x_-+{\frac{2\alpha \hbar}{1+\alpha P_+}}M+2\epsilon\hbar\left(
\frac{\alpha}{1+\alpha P_+}-\frac{\ln(1+\alpha P_+)}{P_+}\right){\frac{P_j}{P_+}}M_{+j},\nonumber\\
x_j^{\cal F}&=&x_j -\epsilon\epsilon_{jk} \hbar\ln(1+\alpha P_+) \frac{1}{P_+}M_{+k}.
\eea
The bullet $\bullet$ denotes $\pm, j$ (all translation generators are undeformed). The summation over repeated indices is understood.

\section{The hermiticity and pseudo-hermiticity condition for twist-deformed generators}

For an operator $\Omega$ the hermiticity condition is $\Omega^\dagger =\Omega$.\par
The pseudohermiticity condition \cite{mos} is 
\bea \label{pseudo}
\Omega^\dagger &=&\eta\Omega\eta^{-1},
\eea 
for some invertible hermitian operator $\eta=\eta^\dagger$. \par
For the jordanian twist the pseudo-hermiticity properties of the deformed generators are the following.\par
In case $I$ (jordanian ${\cal F}$ twist with $\epsilon=0$):
\bea
{P_\bullet^{\cal F}}^\dagger &=& \eta^\lambda P_\bullet^{\cal F} \eta^{-\lambda}, \quad \forall \lambda\in \mathbb{R},\nonumber\\
{M^{\cal F}}^\dagger &=& M^{\cal F}, \quad \quad  ~~~ i.e.  ~\lambda=0,\nonumber\\
{M_{+j}^{\cal F}}^\dagger &=& \eta^{\lambda} M_{+j}^{\cal F} \eta^{-\lambda}, \quad \forall \lambda\in \mathbb{R},\nonumber\\
{M_{-j}^{\cal F}}^\dagger &=& \eta M_{-j}^{\cal F} \eta^{-1}, \quad i.e.~ \lambda=1,\nonumber\\
{N^{\cal F}}^\dagger &=& \eta^\lambda N^{\cal F} \eta^{-\lambda}, \quad\forall \lambda\in \mathbb{R},
\eea
together with
\bea
{x_+^{\cal F}}^\dagger &=& \eta^\lambda {x_+^{\cal F}} \eta^{-\lambda}, \quad \forall\lambda\in {\mathbb{R}},\nonumber\\
{x_-^{\cal F}}^\dagger &=& \eta {x_-^{\cal F}} \eta^{-1}, \quad ~~i.e.~ \lambda=1,\nonumber\\
{x_j^{\cal F}}^\dagger&=& \eta^\lambda {x_j^{\cal F}} \eta^{-\lambda}, \quad \forall\lambda\in {\mathbb{R}},\nonumber\\
\eea
for the hermitian operator $\eta=1+\alpha P_+$.\par
One can observe that, in this case, the subset of hermitian ($\lambda=0$) deformed operators is given by $\{P_\pm^{\cal F}, P_j^{\cal F}, M^{\cal F}, N^{\cal F}, M_{+j}^{\cal F}, x_+^{\cal F}, x_j^{\cal F}, \hbar^{\cal F}\}$, since the operators
$M_{-j}^{\cal F}, x_-^{\cal F}$ are not hermitian. \par
In the case $II$ (jordanian ${\cal F_\tau}$ twist with $\epsilon=0$) we have
\bea\label{pseudoher}
{P_\bullet^{\cal F}}^\dagger &=& \eta^\lambda P_\bullet^{\cal F} \eta^{-\lambda}, \quad \forall \lambda\in \mathbb{R},\nonumber\\
{M^{\cal F}}^\dagger &=& \eta M^{\cal F}\eta^{-1},   ~~~ i.e.  ~\lambda=1,\nonumber\\
{M_{+j}^{\cal F}}^\dagger &=& \eta^{\lambda} M_{+j}^{\cal F} \eta^{-\lambda}, \quad \forall \lambda\in \mathbb{R},\nonumber\\
{N^{\cal F}}^\dagger &=& \eta^\lambda N^{\cal F} \eta^{-\lambda}, \quad\forall \lambda\in \mathbb{R},
\eea
together with
\bea
{x_+^{\cal F}}^\dagger &=& \eta^\lambda {x_+^{\cal F}} \eta^{-\lambda}, \quad \forall\lambda\in {\mathbb{R}},\nonumber\\
{x_j^{\cal F}}^\dagger&=& \eta^\lambda {x_j^{\cal F}} \eta^{-\lambda}, \quad \forall\lambda\in {\mathbb{R}},\nonumber\\
\eea
for the hermitian operator $\eta=\frac{1+\alpha P_+}{1+2\alpha P_+}$.\par
Unlike the other deformed generators,  $M_{-j}^{\cal F}, x_-^{\cal F}$ do not satisfy the pseudo-hermiticity
condition for any choice of $\eta$.\par
It is worth pointing out that, by taking the choice $\lambda=1$, we formally obtain the same set of $P_\pm^{\cal F}, P_j^{\cal F}, M^{\cal F}, N^{\cal F}, M_{+j}^{\cal F}, x_+^{\cal F}, x_j^{\cal F}, \hbar^{\cal F}$ deformed generators as in the previous case. They are now, of course, different operators which satisfy a  pseudo-hermiticity condition. As we shall see, they close a nonlinear algebra.\par
For the extended jordanian twist, cases $III$ and $IV$, one can verify by explicit computation through
a Taylor expansion in the deformation parameter $\alpha$, that most of the deformed generators are neither
hermitian nor pseudo-hermitian.

\section{Deformed algebras as nonlinear algebras}

In the case $I$ (the jordanian ${\cal F}$ twist with $\epsilon=0$) the deformed generators induce nonlinear (at most quadratic) algebras {\textcolor{black}{recovered from their commutators (see the comments in Appendix {\bf B}).}} The basis of deformed generators defining the ${\cal G}_{dfr}$ nonlinear algebra is given
by
\bea
{\cal G}_{dfr}&:& \{\hbar^{\cal F}, x_\pm^{\cal F}, x_j^{\cal F}, P_\pm^{\cal F}, P_j^{\cal F}, M^{\cal F}, N^{\cal F}, M_{\pm j}^{\cal F}\}.
\eea
{\textcolor{black}{The deformed generators are obtained from the ${\cal G}=\{P_\mu, M_{\mu\nu}, x_\mu, \hbar\}$ set of Lie algebra generators after applying the
(\ref{twistdefmap}) mapping.}}
Some of its relevant subalgebras are the deformed Poincar\'e subalgebra ${\cal P}_{dfr}$, with basis of deformed generators given by
\bea
{\cal P}_{drf}&:& \{ P_\pm^{\cal F}, P_j^{\cal F}, M^{\cal F}, N^{\cal F}, M_{\pm j}^{\cal F}\},
\eea
as well as the subalgebra of mutually consistent observables (generators with the same hermiticity/pseudo-hermiticity property).  The ${\cal G}_{obs}$ and the ${\cal P}_{obs}$ subalgebras of observables are respectively given by
\bea
{\cal G}_{obs}&:&\{\hbar^{\cal F}, x_+^{\cal F}, x_j^{\cal F}, P_\pm^{\cal F}, P_j^{\cal F}, M^{\cal F}, N^{\cal F}, M_{+ j}^{\cal F}\}
\eea
and
\bea
{\cal P}_{obs}&:&\{ P_\pm^{\cal F}, P_j^{\cal F}, M^{\cal F}, N^{\cal F}, M_{+ j}^{\cal F}\}.
\eea
The non-vanishing commutation relations of the ${\cal G}_{dfr}$ nonlinear algebra are explicitly given by
\bea\label{nlinal}
\relax [{P}_{\pm}^{\cal F},M^{\cal F}]&= &\pm i P_\pm^{\cal F} \mp i\alpha P_+^{\cal F} P_{\pm}^{\cal F},\nonumber\\
\relax [{P}_{-}^{\cal F},M_{+j}^{\cal F}]&= &2 i P_j^{\cal F} ,\nonumber\\
\relax [{P}_{-}^{\cal F},M_{-j}^{\cal F}]&= &2 i\alpha P_-^{\cal F} P_{j}^{\cal F},\nonumber\\
\relax [{P}_{j}^{\cal F},M_{\pm k}^{\cal F}]&= & i \delta_{jk} P_{\pm}^{\cal F},\nonumber\\
\relax [{M}^{\cal F},M_{\pm j}^{\cal F}]&= &\mp i M_{\pm j}^{\cal F} \pm  i\alpha M_{\pm j}^{\cal F} P_{+}^{\cal F},\nonumber\\
\relax [{N}^{\cal F},M_{\pm j}^{\cal F}]&= & i\epsilon_{jk}  M_{\pm k}^{\cal F},\nonumber\\
\relax [{M}_{+j}^{\cal F},M_{-k}^{\cal F}]&= &-2 i \delta_{jk}M^{\cal F} -2 i
\epsilon_{jk} N^{\cal F}-2i\alpha M_{+j}^{\cal F} P_{k}^{\cal F},\nonumber\\
\relax [{M}_{-j}^{\cal F},M_{-k}^{\cal F}]&= &2 i \alpha ( M_{-j}^{\cal F}P_k^{\cal F} -M_{-k}^{\cal F}P_j^{\cal F}),\nonumber\\
\relax [{P}_{+}^{\cal F},x_-^{\cal F}]&= &2 i\hbar^{\cal F}(1-\alpha P_+^{\cal F}),\nonumber\\
\relax [{P}_{-}^{\cal F},x_{+}^{\cal F}]&= &2 i\hbar^{\cal F},\nonumber\\
\relax [{\cal P}_{-}^{\cal F},x_{-}^{\cal F}]&= &2 i\alpha\hbar^{\cal F} P_-^{\cal F},\nonumber\\
\relax [{\cal P}_{j}^{\cal F},x_{k}^{\cal F}]&= &-i\delta_{jk} \hbar^{\cal F},\nonumber\\
\relax [{M}^{\cal F},x_{\pm}^{\cal F}]&= &\mp  i x_{\pm}^{\cal F} \pm i\alpha x_{\pm}^{\cal F} P_{+}^{\cal F},\nonumber\\
\relax [{N}^{\cal F},x_{j}^{\cal F}]&= &i\epsilon_{jk} x_k^{\cal F},\nonumber\\
\relax [{M}_{+j}^{\cal F},x_{-}^{\cal F}]&= &-2 i x_j^{\cal F} -2 i\alpha \hbar^{\cal F} M_{+j}^{\cal F},
\nonumber\\
\relax [{M}_{-j}^{\cal F},x_{+}^{\cal F}]&= &-2 i x_j^{\cal F} +2 i\alpha x_+^{\cal F} P_{j}^{\cal F},\nonumber\\
\relax [{M}_{-j}^{\cal F},x_{-}^{\cal F}]&= &2 i\alpha (\hbar^{\cal F} M_{-j}^{\cal F} -x_-^{\cal F} P_{j}^{\cal F}),
\nonumber\\
\relax [{M}_{\pm j}^{\cal F},x_{k}^{\cal F}]&= &- i\delta_{jk} x_{\pm}^{\cal F},\nonumber\\
\relax [{x}_{+}^{\cal F},x_{-}^{\cal F}]&= &-2 i\alpha \hbar^{\cal F} x_+^{\cal F}.
\eea
From the above formulas one can check that ${\cal P}_{dfr}$, ${\cal G}_{obs}$ and ${\cal P}_{obs}$ close as
nonlinear subalgebras.\par
It follows from the last formula of the (\ref{nlinal}) algebra that the associated non-commutative space-time  
belongs to the class of Lie-deformed space-times, see e.g. \cite{luwo}, with light-like deformation.\par
In case $II$ (the jordanian ${\cal F}_\tau$ twist) the algebra of the deformed observables,
given by the operators $\hbar^{\cal F}, x_+^{\cal F}, x_j^{\cal F}, P_\pm^{\cal F}, P_j^{\cal F}, M^{\cal F}, N^{\cal F}, M_{+ j}^{\cal F}$ obtained from formulas (\ref{twistdefiione}) and (\ref{twistdefiitwo}) with $\epsilon=0$, also closes
as a nonlinear algebra. In this class of operators the only one which differs from its undeformed counterpart is
$M^{\cal F}$, given by
\bea
M^{\cal F}= ZM &,& Z= \frac{1+2\alpha P_+}{1+\alpha P_+}= \frac{1+2\alpha P_+^{\cal F}}{1+\alpha P_+^{\cal F}}.
\eea It turns out that the commutators with nonlinear
right hand side are the following ones:
\bea
\relax [ M^{\cal F}, P_\pm^{\cal F}]&=& \mp i Z P_\pm^{\cal F},\nonumber\\
\relax [ M^{\cal F}, M_{+j}^{\cal F}]&=& - i Z M_{+j}^{\cal F},\nonumber\\
\relax [ M^{\cal F}, x_+^{\cal F}]&=& - i Z x_+^{\cal F}.
\eea
All the remaining commutators are linear and coincide with the ones (see the Appendix {\bf A}) for the undeformed generators.

\section{Bounded domains and nonlinear additivity in deformed systems}

The second order (mass-term) Casimir of the Poincar\'e algebra
\bea
C_2 &=& P_+P_--({P_j})^2
\eea
remains undeformed under the twist-deformations $I, II, IV$ introduced in Section {\bf 2}. We get
\bea
C_2&=& P_+^{\cal F}P_-^{\cal F} - ({P_j^{\cal F}})^2.
\eea
All ten twisted generators (collectively denoted as $g_I^{\cal F}$, $I=1,\ldots, 10$) entering their respective deformed Poincar\'e algebras, commute with $C_2$:
\bea
\relax [g_I^{\cal F}, C_2]&=&0, \quad \forall g_I^{\cal F}.
\eea

To analyze the physical consequences of the deformation we consider here the simplest setting, namely the case $I$ (the ${\cal F}$ jordanian twist with $\epsilon=0$). We will discuss the properties of the deformed momenta and their nonlinear addition formulas. \par
Let's set, for simplicity, the transverse momenta
$P_j\equiv 0$. For convenience we introduce new variables, defined as
\bea
r= P_+, && l= P_-,\nonumber\\
e = P_0,&&
p= P_1.
\eea
Since $P_\pm =P_0\pm P_1$, we have
\bea
e= \frac{1}{2}(r+l),\quad&&\quad
p= \frac{1}{2}(r-l),
\eea
with $e$ representing the energy of the system.\par
For a massive representation we get, on shell,
\bea
rl= m^2,\quad&&\quad
e^2-p^2 =m^2.
\eea
Let's set, without loss of generality, $m=1$. Therefore
\bea
rl&=& 1,\nonumber\\
e^2-p^2&=&1,\nonumber\\
l&=&\frac{1}{r},\nonumber\\
e&=& \frac{1}{2}(r+\frac{1}{r}),\nonumber\\
p&=& \frac{1}{2}(r-\frac{1}{r}).
\eea
In order to have a positive energy $e$, $r$ should be non-negative. The observables are therefore bounded
in the domains
\bea
\relax r&\in& ]0,+\infty] ,\nonumber\\
\relax l&\in& ]0,+\infty],\nonumber\\
\relax e &\in& [1,+\infty],\nonumber\\
\relax p&\in&[-\infty, +\infty ].
\eea
The rest condition for the $P_1$ momentum corresponds to $p=0$ obtained at $r=1$. For this value the energy is minimal ($e=1$).\par
The twist-deformed variables will be denoted with a bar, the deformation parameter being $\alpha$. We have
\bea
{\overline r} = \frac{r}{1+\alpha r},\quad&&\quad
{\overline l} = l(1+\alpha r).
\eea
The condition
\bea
\alpha&\geq&0
\eea
has to be imposed to avoid singularities.\par
On-shell we have
\bea
{\overline l} &=& \frac{1}{r}(1+\alpha r),\nonumber\\
{\overline e} &=& \frac{1}{2}({\overline r}+{\overline l}) = \frac{1}{2}\left(\frac{r}{1+\alpha r}+\frac{1}{r}(1+\alpha r)\right),\nonumber\\
{\overline p} &=& \frac{1}{2}({\overline r}-{\overline l}) = \frac{1}{2}\left(\frac{r}{1+\alpha r}-\frac{1}{r}(1+\alpha r)\right).
\eea
The rest condition ${\overline p}=0$ for the deformed $P_1$ momentum is obtained
for
\bea
r&=& \frac{1}{1-\alpha}.
\eea
Since $r$ should be positive, one finds the restriction $\alpha<1$. Therefore, the range of the deformation parameter $\alpha$ is
\bea
&0\leq \alpha<1.&
\eea
As a consequence, the domains of the deformed operators are modified with respect to their undeformed 
counterparts. We get
\bea\label{domain}
\relax {\overline r}&\in& ]0,\frac{1}{\alpha}[,\nonumber\\
\relax {\overline l} &\in& ]\alpha,+\infty ],\nonumber\\
\relax {\overline e}&\in& [1,+\infty],\nonumber\\
\relax {\overline p}&\in& [-\infty, \frac{1}{2}(\alpha-\frac{1}{\alpha})[.
\eea

Following \cite{ourtwist4} {\textcolor{black}{(see also the explanations in Appendix {\bf B}) the additive formulas of the deformed momenta are induced by their deformed coproducts.}} In particular, 
the $2$-particle addition formula for the deformed $P_+$ momenta ${\overline r}_1, {\overline r}_2$ of, respectively, the first and the second particle reads, in terms of the undeformed momenta, 
as
\bea
{\overline r}_{1+2} &=& \frac{r_1+r_2}{1+\alpha(r_1+r_2)}.
\eea
Closely expressed in terms of the deformed momenta it is given by
\bea\label{defaddit}
{\overline r}_{1+2} &=&\frac{ {\overline r}_1+{\overline r}_2-2\alpha{\overline r}_1{\overline r}_2}{1-\alpha^2{\overline r}_1{\overline r}_2}.
\eea
The binary operation defined by Eq. (\ref{defaddit}) is a group operation with ${\overline r}=0$ as identity element
and inverse element  given by ${\overline r}^{-1} = -\frac{{\overline r}}{1-2\alpha{\overline r}}$. The associativity is satisfied due to the relation
\bea
{\overline r}_{(1+2)+3}= {\overline r}_{1+(2+3)} &=& \frac{{\overline r}_1+{\overline r}_2+{\overline r}_3-2\alpha({\overline r}_1{\overline r}_2+{\overline r}_2{\overline r}_3+{\overline r}_3{\overline r}_1)+3\alpha^2{\overline r}_1{\overline r}_2{\overline r}_3}{
1-\alpha^2({\overline r}_1{\overline r}_2+{\overline r}_2{\overline r}_3+{\overline r}_3{\overline r}_1)+2\alpha^3{\overline r}_1{\overline r}_2{\overline r}_3}.
\eea
It should be noted, however, that the physical requirement of ${\overline r}$ belonging to the domain in Eq. (\ref{domain}) excludes the identity and the inverse element from the physical values. Thus, on physical states, the addition formula (\ref{defaddit}) only satisfies the properties of a semigroup operation.\par
It is useful to compare the formula (\ref{defaddit}) with the nonlinear addition of velocities in special relativity.
Let us change variables once more and set ${\overline r}_{1,2} =v_{1,2}$, $\alpha=\frac{1}{c}$.\par
In special relativity we have
\bea
v_{1+2,s.r.} &=& \frac{v_1+v_2}{1+\frac{1}{c^2}v_1v_2}.
\eea
In the above jordanian deformation we get
\bea
{v}_{1+2,jd.} &=&\frac{ {v}_1+{v}_2-\frac{2}{c}{v}_1{v}_2}{1-\frac{1}{c^2}{v}_1{v}_2}.
\eea

Both nonlinear addition formulas are symmetric in the $v_1\leftrightarrow v_2$ exchange and associative.

 They can also be defined in the interval $0\leq v_{1,2}\leq c$, so that the nonlinear additive velocities belong to the $[0,c]$ range (in both cases if $v_1=0$, then $v_{1+2}=v_2$
and, if $v_1=c$, $v_{1+2}=c$).\par
The main difference is that in special relativity the formula can be nicely extended to negative velocities
belonging to the $-c\leq v_{1,2}\leq c$ interval. This is not the case for the nonlinear addition formula based on the jordanian deformation.\par
It is worth to point out that an important feature of the nonlinear additivity consists in respecting the physical domain of the variables defining the multi-particle state.\par
For completeness and as an example we present the coproducts for the twist-deformed translation generators $P_\pm^{\cal F}, P_j^{\cal F}$ induced by the case $I$ twist. The nonlinear addition formulas are recovered as a consequence. In principle one should use the twist-deformed coproduct $\Delta^{\cal F}$. However, as recalled in Appendix {\bf B}, the application of the ordinary coproduct $\Delta$ (whose computations are easier)
produces unitarily equivalent results. \par
We set, for convenience,
\bea
&{\overline r}_1= P_+^{\cal F} \otimes {\bf 1}, ~
{\overline r}_2= {\bf 1}\otimes P_+^{\cal F},~
{\overline s}_1= P_-^{\cal F} \otimes {\bf 1}, ~
{\overline s}_2= {\bf 1}\otimes P_-^{\cal F},~  
{\overline t_j,}_1= P_j^{\cal F} \otimes {\bf 1}, ~
{\overline t_j,}_2= {\bf 1}\otimes P_j^{\cal F}. &\nonumber
\eea
The coproducts are
\bea
\Delta (P_+^{\cal F} )&=& \Delta(P_+)\cdot \frac{\bf 1}{ {\bf 1}+\alpha \Delta(P_+)},
\eea
which implies the nonlinear additive formula (\ref{defaddit}),
\bea
\Delta (P_-^{\cal F} )&=& P_-^{\cal F}\otimes {\bf 1} +{\bf 1}\otimes P_-^{\cal F} +\alpha P_-^{\cal F}({\bf 1}-\alpha P_+^{\cal F})\otimes P_+^{\cal F}\frac{\bf 1}{({\bf 1} -\alpha P_+^{\cal F})}+\nonumber\\
&& \alpha P_+^{\cal F}\frac{\bf 1}{({\bf 1} -\alpha P_+^{\cal F})} \otimes P_-^{\cal F}({\bf 1}-\alpha P_+^{\cal F}),
\eea
which implies the nonlinear additive formula
\bea
{\overline s}_{1+2} &=& {\overline s}_1+{\overline s}_2+ \alpha  {\overline s}_1 {\overline r}_2
\frac{1-\alpha {\overline r}_1}{1-\alpha  {\overline r}_2}+\alpha  {\overline r}_1 {\overline s}_2
\frac{1-\alpha {\overline r}_2}{1-\alpha  {\overline r}_1},
\eea
\bea
\Delta (P_j^{\cal F} )&=& P_j^{\cal F}\otimes {\bf 1} +{\bf 1}\otimes P_j^{\cal F},
\eea
which gives, since the transverse momenta are undeformed,
\bea
{\overline t_j,}_{1+2} &=& {\overline t_j,}_1+{\overline t_j,}_2.
\eea

\section{Conclusions}

In this paper we investigated the observational consequences of the light-like deformations of the Poincar\'e algebra induced by the jordanian
and the extended jordanian twists (which both belong to the class of Drinfel'd twist deformations). 
 We used the framework of \cite{ourtwist1}-\cite{ourtwist4} where, in particular, the deformed quantum theory is recovered from deformed twist-generators
belonging to a Universal Enveloping Algebra.  The Hopf algebra structure of the twist-deformation controls the
physical properties of the theory.  The deformed generators induce nonlinear algebras, while the coproduct
implies associative nonlinear additivity of the multi-particle states. In the simplest setting of the jordanian deformation along a light-cone direction, these nonlinearities consistently imply the existence of a maximal light-cone momentum. The undeformed theory is recovered by allowing the maximal light-cone momentum to go to infinity. This situation finds a parallelism, as we noted, with the light velocity as the maximal speed in special relativity.\par
A question that we raised regards the status, as observables of a quantum theory, of the deformed generators.  
It is rewarding that for the jordanian deformation a large subset of the deformed Poincar\'e generators are pseudo-hermitian for the same choice of an invertible hermitian $\eta$ operator, see e.g. formula (\ref{pseudoher}). Therefore, they are observable with respect to the $\eta$-modified inner product \cite{mos}.  This subset of observables close a nonlinear algebra as a consistent deformation of a
Poincar\'e subalgebra.\par
The picture is different for the extended jordanian twist. One can explicitly check, by Taylor-expanding in power series of the deformation parameter, that most of the Poincar\'e deformed generators are neither hermitian nor pseudo-hermitian. It is still an open question the status as observables of the extended jordanian twist-deformed generators. \par
We point out that, so far, no systematic study has been conducted to investigate the physical consequences of all $21$ classes of Poincar\'e twist deformations given in \cite{tol}. The case we are addressing here is particularly relevant because, according to  \cite{tol}, each twist can be obtained as the product of factorized twists. The three-parameter twist entering the second case of Table $1$ is the product of three factorized twists, two of them being abelian, while the third one is our extended jordanian twist. Setting to zero the two extra deformation parameters as done in this paper corresponds to analyze the ``pure" extended jordanian twist. The two extra abelian deformations are easier to be invesitigated.

~\\

{\bf{\Large Appendix A: Poincar\'e algebra in the light-cone basis.}}\\
~\\
We work with the metric $\eta_{\mu\nu}= diag (-1,1,1,1)$ ($\mu,\nu=0,1,2,3$).\par
The Poincar\'e algebra is given by the commutators
\bea\label{poincare}
\relax [P_\mu,P_\nu]&=&0,\nonumber\\
\relax [M_{\mu\nu},P_\rho]&=&i\eta_{\mu\rho}P_\nu-i\eta_{\nu\rho}P_\mu,\nonumber\\
\relax [M_{\mu\nu},M_{\rho\sigma}]&=& i\eta_{\mu\rho}M_{\nu\sigma}-i\eta_{\mu\sigma}M_{\nu\rho}-i\eta_{\nu\rho}M_{\mu\sigma}+i\eta_{\nu\sigma}M_{\mu\rho}.
\eea
In terms of the spacetime coordinates $x^\mu$ ($x_\mu$), with $\mu= 0, j$ and $j=1,2,3$, the Poincar\'e generators can be realized through the positions
\bea
P_0&=& i\hbar\frac{\partial}{\partial x_0},\nonumber\\
P_j&=& -i\hbar\frac{\partial}{\partial x_j},\nonumber\\
M_{0j}&=& -ix_0\frac{\partial}{\partial x_j}-ix_j\frac{\partial}{\partial x_0},\nonumber\\
M_{jk}&=& -ix_j\frac{\partial}{\partial x_k}+ix_k\frac{\partial}{\partial x_j}.
\eea

$P_\mu, M_{\mu\nu}$, together with the coordinates $x_\mu$ and the central charge $\hbar$ close a Lie algebra ${\cal G}=\{P_\mu, M_{\mu\nu}, x_\mu, \hbar\}$ whose remaining non-vanishing commutators are given by
\bea
\relax [P_0,x_0] &=& i\hbar,\nonumber\\
\relax [P_j, x_k]&=& -i\delta_{jk}\hbar,\nonumber\\
\relax [M_{0j}, x_0] &=& -i x_j, \nonumber\\
\relax [ M_{0j},x_k]&=& -ix_0\delta_{jk},\nonumber\\
\relax [ M_{jk}, x_r]&=& -ix_j\delta_{kr}+ix_k\delta_{jr}.
\eea
The light-cone coordinates mix the time-direction $x_0$ and the space direction $x_1$;
the remaining space coordinates are the transverse directions. The light-cone generators are labeled by the indices $\pm$ and $j=1,2$. The latter is used for the transverse coordinates.\par
In the light-cone basis the generators are expressed through the positions
\bea
{\overline x}_\pm = x_0\pm x_1, &\quad&  {\overline x}_j = x_{j+1}
\eea
and
\bea
{\overline P}_\pm =P_0\pm P_1, &\quad& {\overline P_j }= P_{j+1},\nonumber\\
{\overline M}&:=&{\overline M}_{+-} = M_{01},\nonumber\\
{\overline M}_{\pm j} &=& M_{0,j+1}\pm M_{1,j+1},\nonumber\\
{\overline N} &=& M_{23}.
\eea
In the above formulas relating the two presentations of space-time coordinates and Poincar\'e generators, for reason of clarity, the Poincar\'e generators and space-time coordinates expressed
in the light-cone variables are overlined. Since however in the main text we only work with light-cone quantities and no confusion will arise, for simplicity, these quantities will not be overlined. \par
In the light-cone basis the non-vanishing commutators of the Poincar\'e algebra read as follows
\bea
\relax [{M}, {P}_\pm]&=& \mp i{P}_\pm,\nonumber\\
\relax [{M}, {M}_{\pm j}]&=& \mp i{M}_{\pm j},\nonumber\\
\relax [{M}_{\pm j}, {P}_\mp]&=& -2 i{P}_j,\nonumber\\
\relax [{M}_{\pm j}, {P}_k]&=& -i\delta_{jk}P_\pm,\nonumber\\
\relax [{M}_{+j}, {M}_{-k}]&=& -2i\delta_{jk} {M}-2i\epsilon_{jk}{N},\nonumber\\
\relax [{ M}_{\pm j}, { N}]&=& -i \epsilon_{jk} { M}_{\pm k},\nonumber\\
\relax [{N}, {P}_{ j}]&=& i\epsilon_{jk} {P}_k
\eea
(the constant antisymmetric tensor is $\epsilon_{12}=-\epsilon_{21}=1$).\par
The non-vanishing commutators of the Poincar\'e generators with the light-cone coordinates ${ x}_\pm, x_j$ are
\bea
\relax [P_\pm, x_\mp]&=& 2i\hbar,\nonumber\\
\relax [P_j,x_k] &=& -i \delta_{jk}\hbar,\nonumber\\
\relax [ M, x_\pm ] &=& \mp i x_\pm,\nonumber\\
\relax [M_{\pm j}, x_\mp]&=& -2ix_j,\nonumber\\
\relax [ M_{\pm j}, x_k]&=& -i x_\pm \delta_{jk},\nonumber\\
\relax [N, x_j] &=& i\epsilon_{jk}x_k.
\eea

{\bf{\Large Appendix B: comments on the Drinfel'd twist deformation of Hopf algebras.}}\\
~\\

In this paper we investigated Drinfel'd twist deformations of Hopf algebras. We summarize here, for self-consistency, the main features of our approach. Our framework has been detailed before, e.g. in reference \cite{ourtwist2}. It is based on the construction discussed in \cite{asc2}.\par
We point out at first that the cocycles (\ref{unflippedtwist}) and (\ref{flippedtwist}) 
are counitary 2-cocycles satisfying\\
i) $({\bf 1}\otimes {\cal F})(id\otimes \Delta ){\cal F}= ({\cal F}\otimes {\bf 1}) (\Delta\otimes id) {\cal F}$,\\
ii) $(\varepsilon\otimes  id) {\cal F} ={\bf 1} (id\otimes \varepsilon){\cal F}$,\\
iii) ${\cal F} = {\bf 1}\otimes {\bf 1} +{\cal O}(\alpha)$.\\
Any Universal Enveloping Lie Algebra ${\cal U}({\cal G}_{Lie})=A$ over, let's say, the complex field ${\mathbb C}$, is a Hopf algebra
endowed with multiplication $m$, unit $\eta$, coproduct $\Delta$, counit $\varepsilon$, antipode $S$. A twist-deformation satisfying properties i,ii,iii), see \cite{{asc2},{ourtwist2}}, maps the original Hopf algebra into a new Hopf algebra with structures and costructures  given by \cite{dri}
\bea
(A,m, \eta, \Delta, \varepsilon, S)&\rightarrow& (A,m ,\eta,\Delta^{\cal F},\varepsilon, S^{\cal F}),
\eea
where, $\forall a\in A$,
\bea
\Delta^{\cal F} &=&{\cal F}\Delta(a) {\cal F}^{-1},\nonumber \\
S^{\cal F} &=&\chi S\chi^{-1} 
\eea
for $\chi = m((id\otimes S){\cal F})$.\\
For simplicity the new deformed Hopf algebra
is denoted as $A^{\cal F}$. As a vector space $A^{\cal F}= A$ is identified with the original algebra $A$.
As a vector space $A={\cal U}({\cal G}_{Lie})$ can be spanned both by the original Lie algebra generators 
$g_i\in{\cal G}_{Lie}$ or by its twist-deformed generators obtained from (\ref{twistdefmap}), namely
\bea\label{twistdefmap2}
g_i^{\cal F} &=& {\overline f}^\beta(g_i) {\overline f}_\beta, \quad ({\cal F}^{-1}={\overline f}^\beta\otimes
{\overline f}_\beta).
\eea 
This formula was derived in \cite{asc2} on the basis of the \cite{woro} construction.\par
In this paper, following the above prescriptions, we constructed the twist-deformed algebras 
 ${\cal U}^{\cal F}({\cal G})$, ${\cal U}^{\cal F}({\cal P})$, respectively obtained from  
the Lie algebra ${\cal G}=\{P_\mu, M_{\mu\nu}, x_\mu, \hbar\}$
and its  ${\cal P}=\{P_\mu, M_{\mu\nu}\}$ Poincar\'e subalgebra. They are both Hopf algebras.\\
Following the construction of \cite{{ourtwist1},{ourtwist2},{ourtwist4}}, once the twist-deformed generators are
expressed as elements of the Universal Enveloping algebra, relevant  information about the twist-deformed Hopf algebra is encoded in their commutators.\par
To obtain the nonlinear addition formulas of Section {\bf 6} one needs to use a co-structure, given by the twisted coproduct 
$\Delta^{\cal F}(g^{\cal F})$.  There is an important remark; for practical purposes it is more convenient to compute just $\Delta(g^{\cal F})$, see the explanation given in ref. [4]. Indeed, when applied to a Hilbert space, $F\equiv {\cal F}$ becomes a unitary operator, so that ${\widehat {\Delta^{\cal F}(g^{\cal F})}}= F{\widehat{\Delta(g^{\cal F})}}F^{-1}$. Therefore, the operators ${\widehat {\Delta^{\cal F}(g^{\cal F})}}$ and ${\widehat{\Delta(g^{\cal F})}}$ are unitarily equivalent. The non-linearity of the addition formulas is nevertheless guaranteed by the fact that the deformed generators from equation (\ref{twistdefmap2}) are not unitarily equivalent to the undeformed generators. \\
~\\
~\\
\par {\Large{\bf Acknowledgments}}
{}~\par{}~\par
The work received support from CNPq under PQ Grant No. 308095/2017-0.

\end{document}